\documentclass[twocolumn,amssymb,amsmath,floats,showpacs,superscriptaddress,pre,floatfix,longbibliography]{revtex4-1}

\usepackage{amsmath,amssymb,bm}
\usepackage{graphics}
\usepackage{epsfig}
\usepackage{graphicx}
\usepackage{comment}

\newcommand{\avg}[1]{\langle{#1}\rangle}

\begin{document}

\title{Correlated Edge Overlaps in Multiplex Networks}

\author{Gareth J. Baxter}
\affiliation{
Department of Physics $\&$ I3N, University of Aveiro,
3810-193 Aveiro, Portugal}

\author{Ginestra Bianconi}
\affiliation{
School of Mathematical Sciences, Queen Mary University of London, London, E1~4NS, United Kingdom}

\author{Rui~A.~da Costa}
\affiliation{
Department of Physics $\&$ I3N, University of Aveiro,
3810-193 Aveiro, Portugal}
\author{Sergey N. Dorogovtsev}
\affiliation{
Department of Physics $\&$ I3N, University of Aveiro,
3810-193 Aveiro, Portugal}
\affiliation{A.~F.~Ioffe Physico-Technical Institute, 194021 St.~Petersburg, Russia}
\author{Jos\'e F. F. Mendes}
\affiliation{
Department of Physics $\&$ I3N, University of Aveiro,
3810-193 Aveiro, Portugal}

\begin{abstract}
We develop the theory of sparse multiplex networks with partially overlapping links based on their local tree-likeness. This theory 
enables us to find the giant mutually connected component 
in a two-layer multiplex network with 
arbitrary correlations between connections of different types. 
We find that 
correlations between the overlapping and non-overlapping links markedly 
change the phase diagram of the system, leading to multiple hybrid
phase transitions.
For assortative correlations we observe recurrent hybrid phase transitions. 
\end{abstract}


\maketitle

\section{Introduction}\label{Introduction}

Most real networks are not independent but must be treated as sets of interdependent networks (layers) \cite{boccaletti2014structure, kivela2014multilayer}. 
One of the simplest models of complexes of this kind is a multiplex network.
Each layer 
contains the same nodes, but connected by
links specific to that layer. In other words, a multiplex network is a
graph with nodes of one type connected by links of multiple types
(colors). A natural generalization of percolation on a single
network---giant connected component---to multiplex networks is the
giant mutually connected component (mutual component). 
It is defined by the rule that for every pair of nodes in
the mutual component, there must be a path between them in each layer
(which remains within the mutual component).
Under this definition of percolation, a discontinuous  hybrid
transition occurs in sparse multiplex networks
\cite{son2012percolation, baxter2012avalanche}. 

In real networks, physical or other constraints mean that edges from different
layers are likely to be co-located. To cater for this possibility, the
multiplex concept has been further generalised to consider the case
that two nodes may be connected by more than one color of edge \cite{cellai2013percolation, hu2013percolation, min2015link, bianconi2013statistical}
with nonvanishing probability. The simplest example is a two-layer multiplex network.
In this type of network two nodes $i$ and $j$ can be connected in
three different ways: by an edge only in layer $1$, and edge only on
layer $2$, or by edges in both layers, which we will call an overlapping edge.

A message passing approach
 was proposed in Ref. \cite{cellai2013percolation} to characterize
 the giant mutually connected component of
multiplex networks with overlap of the links, but it was later found
\cite{min2015link} that the algorithm characterizes instead a distinct
directed percolation problem for multiplex networks.
Another recent work has proposed a more complex iterative scheme
requiring an intermediate remapping of the network
\cite{hu2013percolation}. This model agrees with  numerical
simulations of the mutually connected component of multiplex networks
with overlap \cite{min2015link}.

Here we consider the more general problem in a two layered multiplex network, in
which we allow arbitrary correlations between the degrees with respect
to the three types of connection. We exploit the locally tree-like structure of infinite sparse random networks to directly write strict self-consistency equations which allow the solution of the problem.

Note the following difference from the problem without overlapped
edges.
A cluster of nodes connected by overlapped edges belongs to the giant
mutual component if at least one node of the cluster is connected to
this component in each layer, even if these nodes are different.
In Ref.~\cite{hu2013percolation}
the calculation was done
by compressing the overlapped clusters
into ``supernodes'', and then considering non-overlapping multiplex
percolation on the resulting network. This requires a rather arduous
process of
finding both mass and degree distributions for these supernodes, and then
incorporating a separate generating function for each size of
overlapped clusters. In Refs.~\cite{hu2013percolation,min2015link} the calculation is done under the assumption overlapped and non-overlapped degrees are uncorrelated. Consideration of degree correlations using this method, while in principle possible, would require modification of a significant step in this calculation.
Here we show that the calculation can in fact
be done straightforwardly in the usual self-consistency equation fashion, making for a much simpler and more direct calculation. Furthermore, no assumptions about correlations need to be made, so arbitrary correlations among the three connection types can be examined without modification of the method.

We are therefore able to confirm the results of
Ref.~\cite{hu2013percolation}, but using a far simpler calculation,
and then generalize them to more complex and interesting situations.
We use our equations to examine the effect of correlations between
overlapped and non-overlapped edge placement. We find that  correlations
qualitatively change the phase diagram, with the giant mutually
connected component emerging through consecutive hybrid 
transitions. Remarkably, in the particular case of assortative
correlations one of these transitions can be recurrent.

\section{Model and equations}\label{equations}

\begin{figure*}
\begin{center}
\scalebox{0.6}{\includegraphics[angle=0]{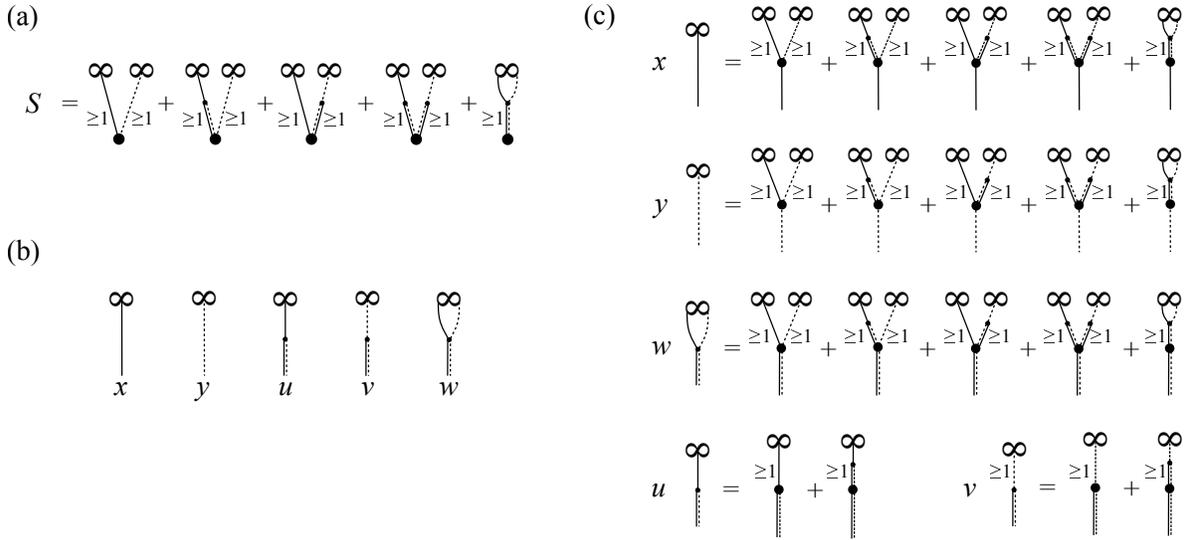}}
\end{center}
\caption{
(a)~Graphical representation of the expression for the size of the mutual component $S$, Eq.~(\ref{10}), and (c)~of the equations for the five probabilities, Eq.~(\ref{20}), 
using notations (b) for five probabilities. 
 } 
 \label{f2}
\end{figure*}

We consider a generalized configuration model for sparse multiplex
networks in the infinite size limit, in which each node has
three degrees $q_1$, $q_2$, $\tilde{q}$, being, respectively, the
number of connections only in layer $1$, only in layer $2$, and the number of
overlapping connections. The network is then defined by the joint
degree distribution
$P(q_1,q_2,\tilde{q})$. 
If a node is connected to the giant mutually connected component in
both layers, then it too belongs to the giant mutually connected
component.
 A single overlapping edge is sufficient to provide this connection.
 As noted above,  one must therefore carefully consider overlapped
clusters when making percolation calculations.
Alternatively at least one single edge of each type is  needed to provide this
connection. 
These considerations lead to the following expression for the relative size $S$ of
the giant mutually connected component, represented graphically in Fig.~\ref{f2}(a):
\begin{widetext}
\begin{align}
S &= \sum_{q_1,q_2,\tilde{q}}P(q_1,q_2,\tilde{q})
\left\{\left[1-(1{-}x)^{q_1}\right] \left[1-(1{-}y)^{q_2}\right] (1{-}u{-}v{-}w)^{\tilde{q}}+\left[1-(1{-}x)^{q_1}\right] \left[(1{-}u{-}w)^{\tilde{q}}-(1{-}u{-}v{-}w)^{\tilde{q}}\right] \right.
\nonumber 
\\
& \left.+ [1{-}(1{-}y)^{q_2}]\left[(1{-}v{-}w)^{\tilde{q}}-(1{-}u{-}v{-}w)^{\tilde{q}}\right]
+ \left[(1{-}w)^{\tilde{q}}-(1{-}w{-}u)^{\tilde{q}}-(1{-}w{-}v)^{\tilde{q}}+(1{-}w{-}u{-}v)^{\tilde{q}}\right]+\left[1{-}(1{-}w)^{\tilde{q}}\right]\right\}
\nonumber 
\\[5pt]
&= 1-\sum_{q_1,q_2,\tilde{q}}P(q_1,q_2,\tilde{q})\left[ (1{-}x)^{q_1}(1{-}w{-}u)^{\tilde{q}} + (1{-}y)^{q_2}(1{-}w{-}v)^{\tilde{q}} - (1{-}x)^{q_1}(1{-}y)^{q_2}(1{-}w{-}u{-}v)^{\tilde{q}}\right]
.
\label{10}
\end{align}
\end{widetext}
We define $x$ to be the probability that, on following an arbitrary
edge in layer $1$, we encounter a node belonging to the giant mutual
component, and $y$ as the corresponding probability on following an
edge in layer $2$.  
For overlapping edges, we must consider three
probabilities. First, $u$ is the probability that, on following an
overlapped edge, we
encounter a node with at least one other connection to the giant
mutual component by an edge in layer $1$, and $v$ is the 
probability that the node reached has a connection to the
giant mutual component in layer $2$.
Finally $w$ is the probability that if we follow an arbitrary
overlapped edge we reach a node which has connections to the giant
mutual component in both layer $1$ and layer $2$ (not overlapped).
These probabilities are represented graphically in Fig.~\ref{f2}(b).
They obey the following self-consistency equations,
represented graphically in Fig.~\ref{f2}(c):
\begin{widetext}
\begin{eqnarray}
x&=&1-\sum_{q_1,q_2,\tilde{q}} \frac{q_1}{\avg{q_1}} P(q_1,q_2,\tilde{q})\left[ (1{-}x)^{q_1-1}(1{-}u{-}w)^{\tilde{q}} + (1{-}y)^{q_2}(1{-}v{-}w)^{\tilde{q}} - (1{-}x)^{q_1-1}(1{-}y)^{q_2}(1{-}u{-}v{-}w)^{\tilde{q}}\right]
,
\nonumber 
\\[5pt]
y&=&1-\sum_{q_1,q_2,\tilde{q}} \frac{q_2}{\avg{q_2}} P(q_1,q_2,\tilde{q})\left[ (1{-}x)^{q_1}(1{-}u{-}w)^{\tilde{q}} + (1{-}y)^{q_2-1}(1{-}v{-}w)^{\tilde{q}} - (1{-}x)^{q_1}(1{-}y)^{q_2-1}(1{-}u{-}v{-}w)^{\tilde{q}}\right]
,
\nonumber 
\\[5pt]
u&=&\sum_{q_1,q_2,\tilde{q}}\frac{\tilde{q}}{\avg{\tilde{q}}}P(q_1,q_2,\tilde{q}) (1{-}y)^{q_2} \left[
(1{-}v{-}w)^{\tilde{q}-1}  -(1{-}x)^{q_1}(1{-}u{-}v{-}w)^{\tilde{q}-1}\right]
,
\nonumber 
\\[5pt]
v&=&\sum_{q_1,q_2,\tilde{q}}\frac{\tilde{q}}{\avg{\tilde{q}}}P(q_1,q_2,\tilde{q}) (1{-}x)^{q_1} \left[
(1{-}u{-}w)^{\tilde{q}-1}  -(1{-}y)^{q_2}(1{-}u{-}v{-}w)^{\tilde{q}-1}\right]
,
\nonumber 
\\[5pt]
w&=&1-\sum_{q_1,q_2,\tilde{q}} \frac{\tilde{q}}{\avg{\tilde{q}}} P(q_1,q_2,\tilde{q})\left[ (1{-}x)^{q_1}(1{-}u{-}w)^{\tilde{q}-1} + (1{-}y)^{q_2}(1{-}v{-}w)^{\tilde{q}-1} - (1{-}x)^{q_1}(1{-}y)^{q_2}(1{-}u{-}v{-}w)^{\tilde{q}-1}\right]
.
\label{20}
\end{eqnarray}
\end{widetext}
Solution of Eqs. (\ref{20}) and then substitution into Eq. (\ref{10})
allows one to find the size of the giant mutually connected component
for networks with arbitrary intra- and inter- layer degree
correlations.

If the joint degree distribution is symmetric with respect to the two
layers, i.e.  
$P(q,q',\tilde{q})=P(q',q,\tilde{q})$, this system is reduced
to three equations for $x=y$, $u=v$, and $w$.
We now demonstrate the solution of the system of Eqs. (\ref{20}) in several
representative cases.

\section{Uncorrelated case}\label{uncorrelated}

We first show that previous results for the uncorrelated case can 
straightforwardly be reproduced by our method.
If correlations are absent, and $P(q_1,q_2,\tilde{q}) = {\cal
  P}(q_1,c){\cal P}(q_2,c){\cal P}(\tilde{q},\tilde{c})$, where ${\cal
  P}(q,c)$ is a Poisson distribution with mean $c$, then
$S=x=y=w$, $u=v$, and we arrive at a system of two equations:
\begin{eqnarray}
&&
x = 1 - e^{-cx} e^{-\tilde{c}(x+u)} [2-e^{-(cx+\tilde{c}u)}]
,
\nonumber
\\[5pt]
&&
u = e^{-cx} e^{-\tilde{c}(x+u)} [1-e^{-(cx+\tilde{c}u)}]
.
\label{30}
\end{eqnarray}
The solution of this system readily gives $S(c,\tilde{c})$, the size
of the  giant mutually connected component as a function of the mean
intra-layer degree $c$ and inter-layer degree $\tilde{c}$. The giant mutual component appears with a discontinuous hybrid phase transition. The phase diagram is 
shown in Fig.~\ref{f4}(a).
These results agrees perfectly with the theoretical and numerical
results presented in \cite{hu2013percolation,min2015link}.

\section{Correlations between inter- and intralayer degrees}\label{correlated}

An advantage of out method is that these results may be extended to consider degree correlations without much more difficulty.
We explore the effect of degree correlations by considering random
networks where the average number of overlapping links
$\tilde{c}=f(q_1+q_2)$ is a function of the number of single edges
$q_1+q_2$ for a given node, thus:
\begin{equation}
P(q_1,q_2,\tilde{q})={\cal P}(q_1,c){\cal P}(q_2,c){\cal
  P}[\tilde{q},f(q_1+q_2)].
\label{40}
\end{equation} 
Different forms of the function $f(q_1+q_2)$
allow different types of correlations to be examined.

\subsection{Assortative mixing}\label{assortative}

\begin{figure*}
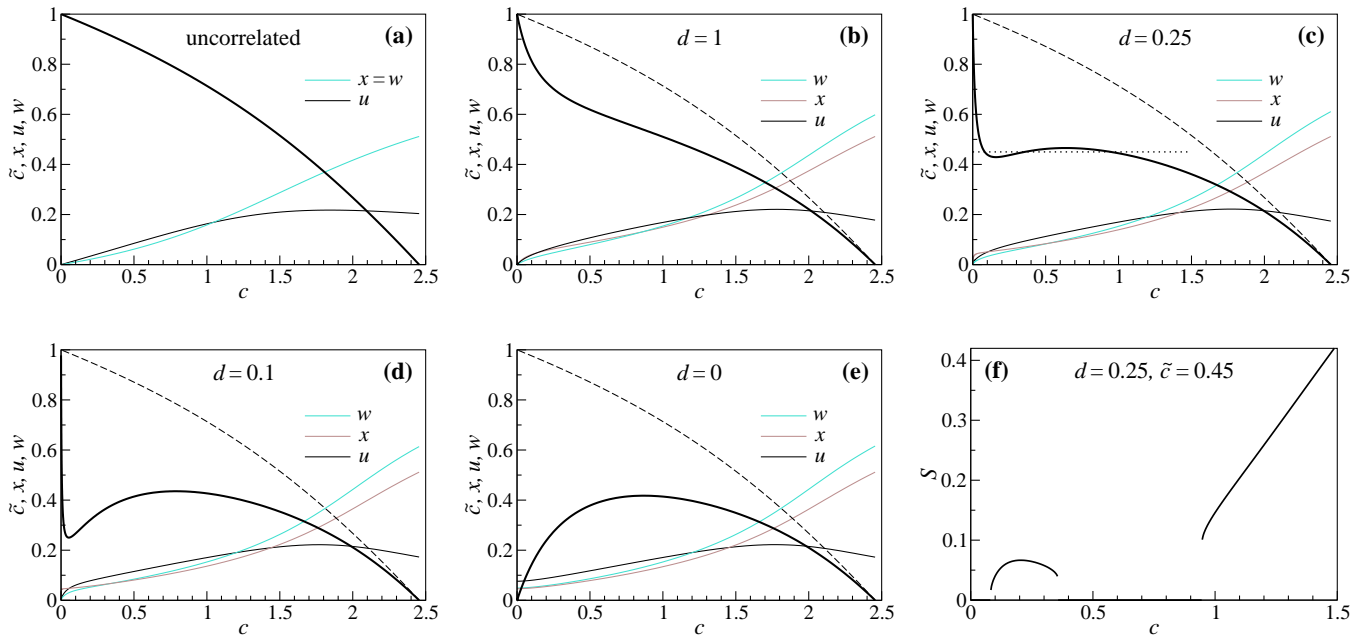

\begin{center}
\includegraphics[width=0.32\textwidth]{fig3a}
\ \ 
\includegraphics[width=0.32\textwidth]{fig3b}
\ \ 
\includegraphics[width=0.32\textwidth]{fig3c}
\\[15pt]
\includegraphics[width=0.32\textwidth]{fig3d}
\ \ 
\includegraphics[width=0.32\textwidth]{fig3e}
\ \ 
\includegraphics[width=0.32\textwidth]{fig3f}
\end{center}
\caption{(Color online) Phase diagrams in the $\tilde{c}$ vs. $c$ plane for the
  uncorrelated and assortatively correlated cases. The appearance of
  the giant mutually connected component with a discontinuous hybrid
  transition is shown as the heavy black line.  
  (a) Symmetric uncorrelated joint degree distribution.  (b)--(e)
  Assortative correlations of Eq.~(\ref{50}) for  
$d=1$, $0.25$, $0.1$, and $0$, respectively. 
For comparison, the dashed line shows the boundary between phases in
the uncorrelated case.
The plots also show the values of the probabilities $x=y$, $u=v$, and
$w$ immediately above the discontinuous transition.
(f) Multiple transitions of $S$ for $d=0.25$, fixed $\tilde{c}=0.45$
and varying $c$, i.e. along the dotted line of panel (c).
}
\label{f4}
\end{figure*}

As a convenient example, we consider assortative correlations using 
a symmetric joint degree distribution 
\begin{equation}
P(q_1,q_2,\tilde{q})={\cal P}(q_1,c){\cal P}(q_2,c){\cal P}[\tilde{q},A(d+q_1+q_2)].
\label{50}
\end{equation}
The parameter
$d$ controls the correlations, with $d\to0$ corresponding to perfect assortativity, and
$d\to\infty$ corresponding to
the uncorrelated case.
The coefficient $A$ normalises the function to maintain the mean
degree $\tilde{c}$ required, $A = \tilde{c}/(d + 2c)$.

Replacing the distribution $P(q_1,q_2,\tilde{q})$ in Eqs.~(\ref{20})
with Eq.~(\ref{50}) we arrive at the system of equations:
\begin{equation}\label{52}
\begin{split}
  & x = 1 - 2 \exp
  \{ {-} 2c {-} A (1 {+} d) (u {+} w)  {-} c(x{-}2)e^{-A(u + w)}
  \}\\
  & \quad +\exp
  \{ {-}  2c  {-} A (1 {+} d) (2u {+} w) {-} 2 c (x{-}1) e^{-A(2u + w
    )}
  \}, \\[5pt]
& u =  \frac{1 - w}{2} - \frac{A}{2\tilde{c}}  \left\{ 2c(1-x)+de^{A (2u + w)} \right\}\\
  & \quad \times \exp
  \{ {-}  2c  {-} A (1 {+} d) (2u {+} w) {-} 2 c (x{-}1) e^{-A(2u + w
    )}
  \},\\[5pt]
& w = 1 - u -  \frac{A}{\tilde{c}} \left\{ c(2-x)+d e^{A(u + w)} \right\}\\
  & \quad \times \exp
  \{ {-} 2c {-} A (1 {+} d) (u {+} w)  {-} c (x{-}2) e^{-A(u + w)}
  \}.
\end{split}
\end{equation}
Then the expression for the relative size of the mutual component $S$
is obtained by substituting Eq.~(5) in  Eqs,~(\ref{20}):
\begin{eqnarray}
S =&& 1 - 2 \exp\left\{ {-} 2c {-} A  d (u {+} w)  {-} c(x{-}2)e^{-A(u + w)} \right\}\ \ \ \ 
\nonumber 
\\[5pt]
&&+ \exp\left\{ {-}  2c  {-} A d (2u {+} w) {-} 2 c (x{-}1) e^{-A(2u + w )} \right\}
.
\label{ss2}
\end{eqnarray}

Figure \ref{f4} shows phase diagrams for different values of the constant $d$.
As the assortativity becomes stronger ($d$ decreases)
the giant mutual component appears at smaller values of $\tilde{c}$, 
except at the endpoints. 
The right-hand end point is at $(c,\tilde{c})=(2.4554...,0)$ (in agreement with the
critical point without overlaps found in \cite{buldyrev2010catastrophic}).

As $d$ approaches $0$ the critical line approaches that found for
$d=0$ (panel (e)),
however the
left-hand end point remains at $(c,\tilde{c})=(0,1)$ for all $d>0$,
and jumps to $(0,0)$ at $d=0$. 
For small $d$ the phase boundary is non-monotonic with respect to $c$, meaning that multiple hybrid transitions may be encountered when following a straight trajectory across the phase plane, as demonstrated in panel (f), which shows the variation of the size $S$ of the mutual component along the dotted line in panel (c). Notice the recurrent hybrid transition after which $S$ returns to zero.

Multiple hybrid transitions have
not previously been observed in this type of system,
although multiple transitions have been noted in networks of networks
\cite{bianconi2014multiple} and several other network
percolation problems \cite{nagler2012continuous,chen2013phase,chen2013unstable,colomer2014double, hackett2016bond}.

\subsection{Disassortative mixing}\label{disassortative}

\begin{figure*}
\begin{center}
\includegraphics[width=0.32\textwidth]{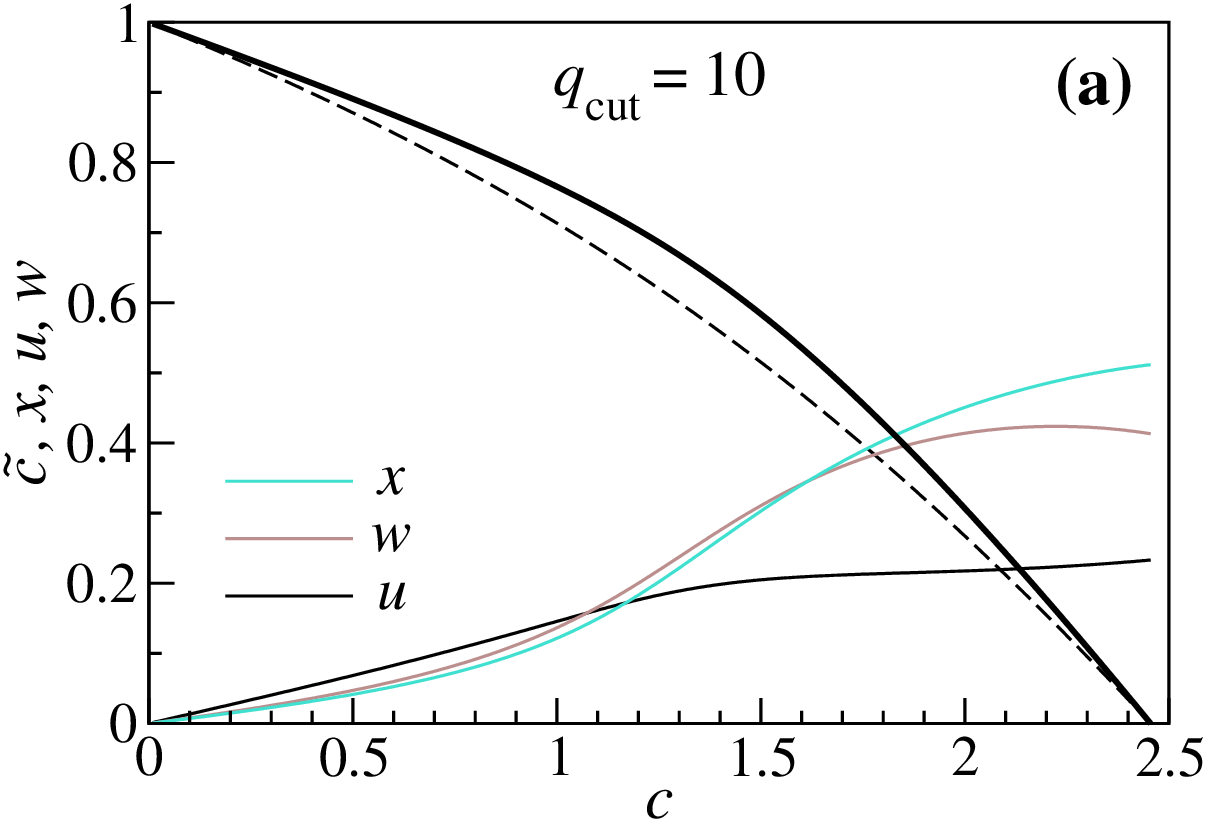}
\ \ 
\includegraphics[width=0.32\textwidth]{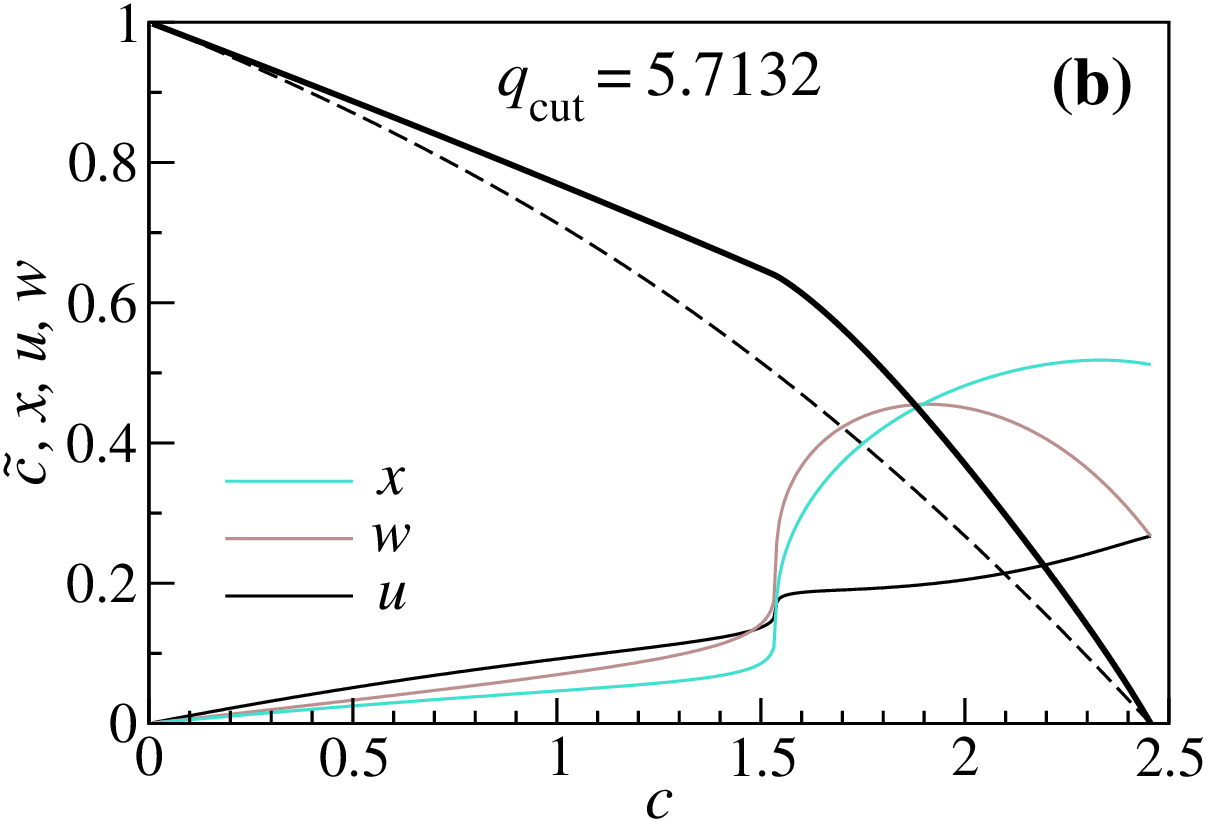}
\ \ 
\includegraphics[width=0.32\textwidth]{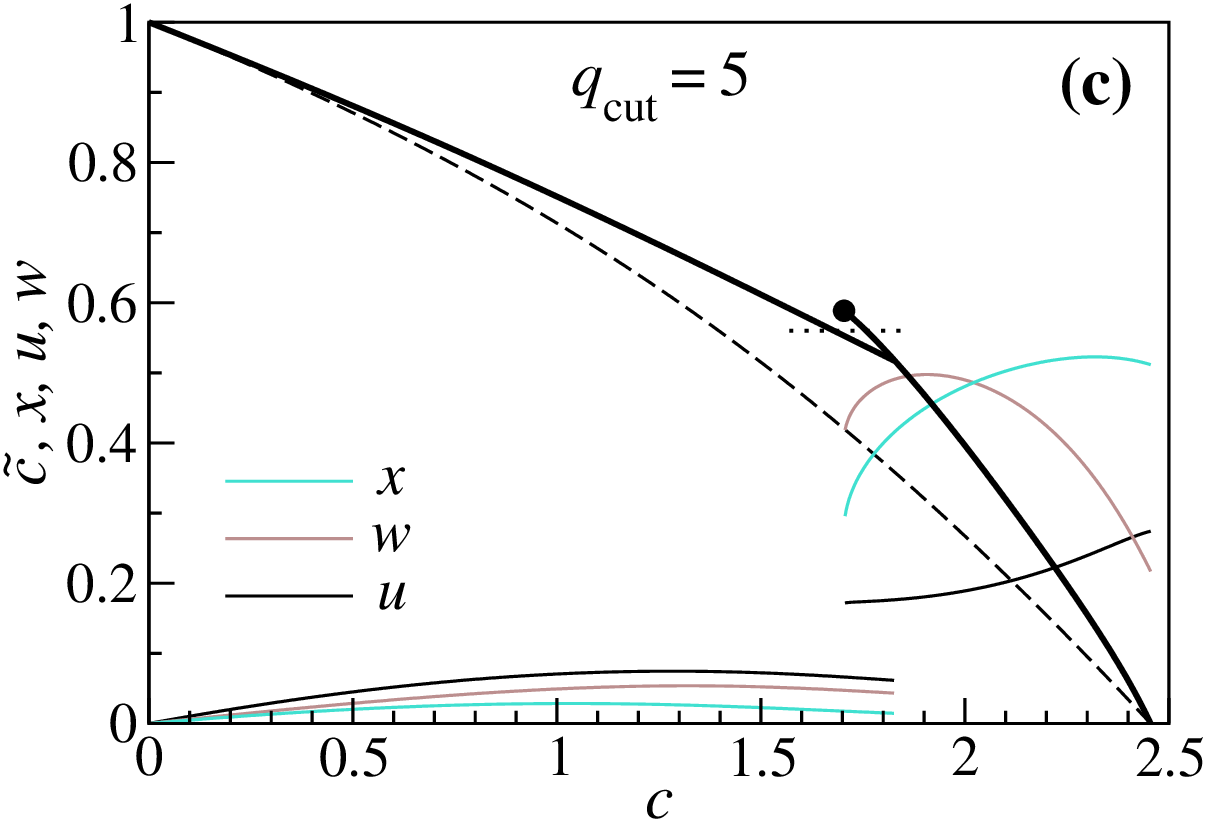}
\\[15pt]
\includegraphics[width=0.32\textwidth]{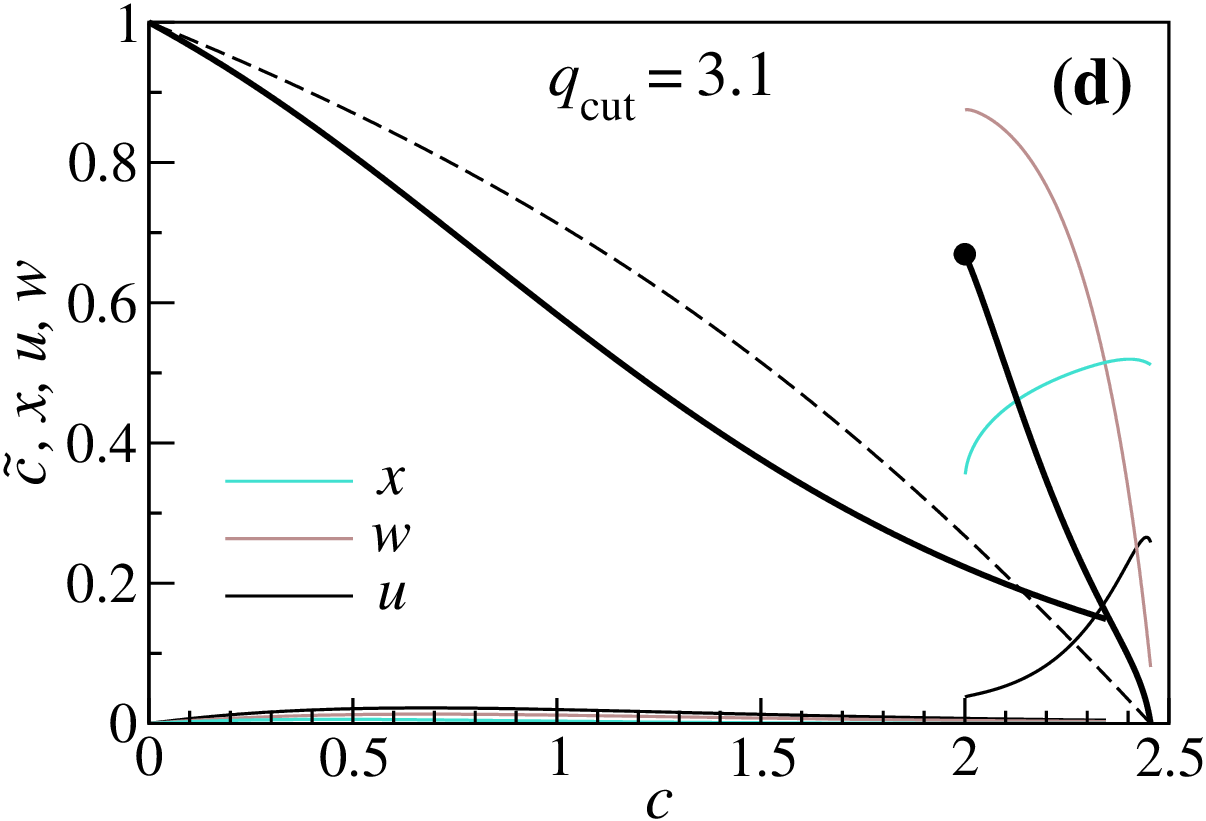}
\ \ 
\includegraphics[width=0.32\textwidth]{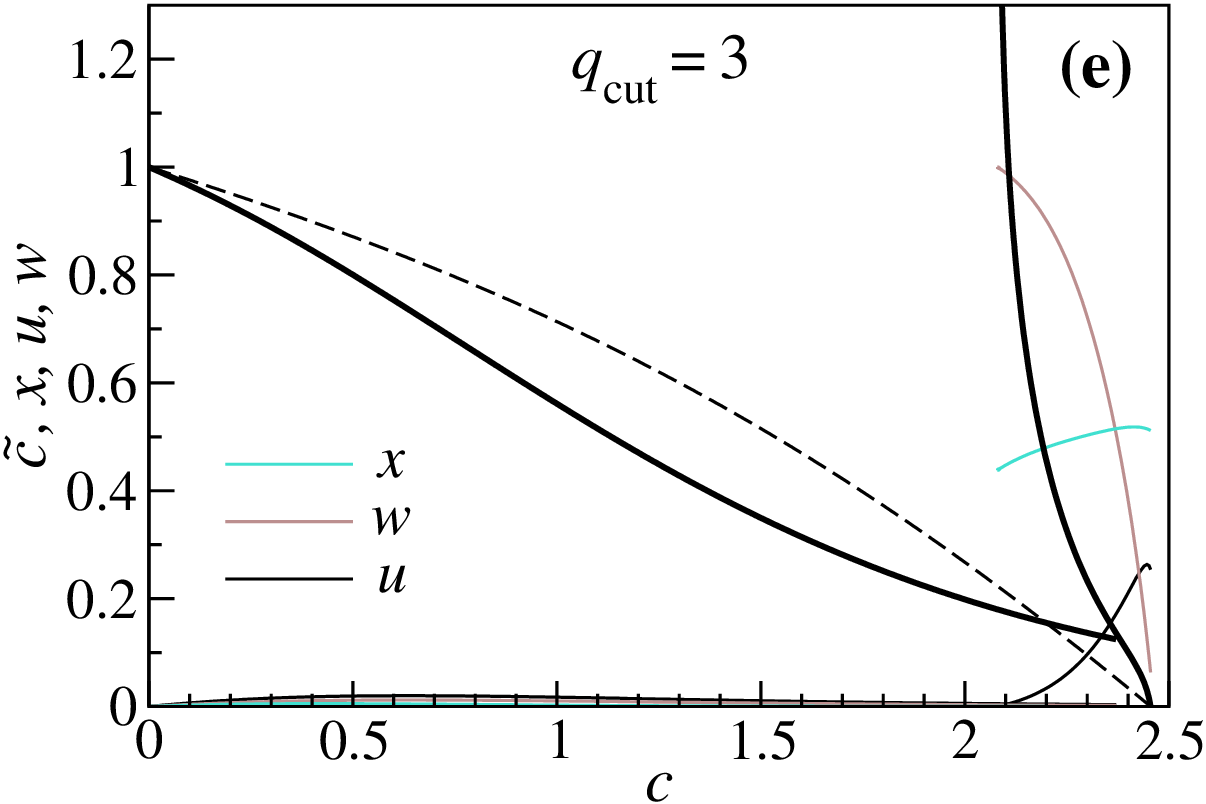}
\ \ 
\includegraphics[width=0.32\textwidth]{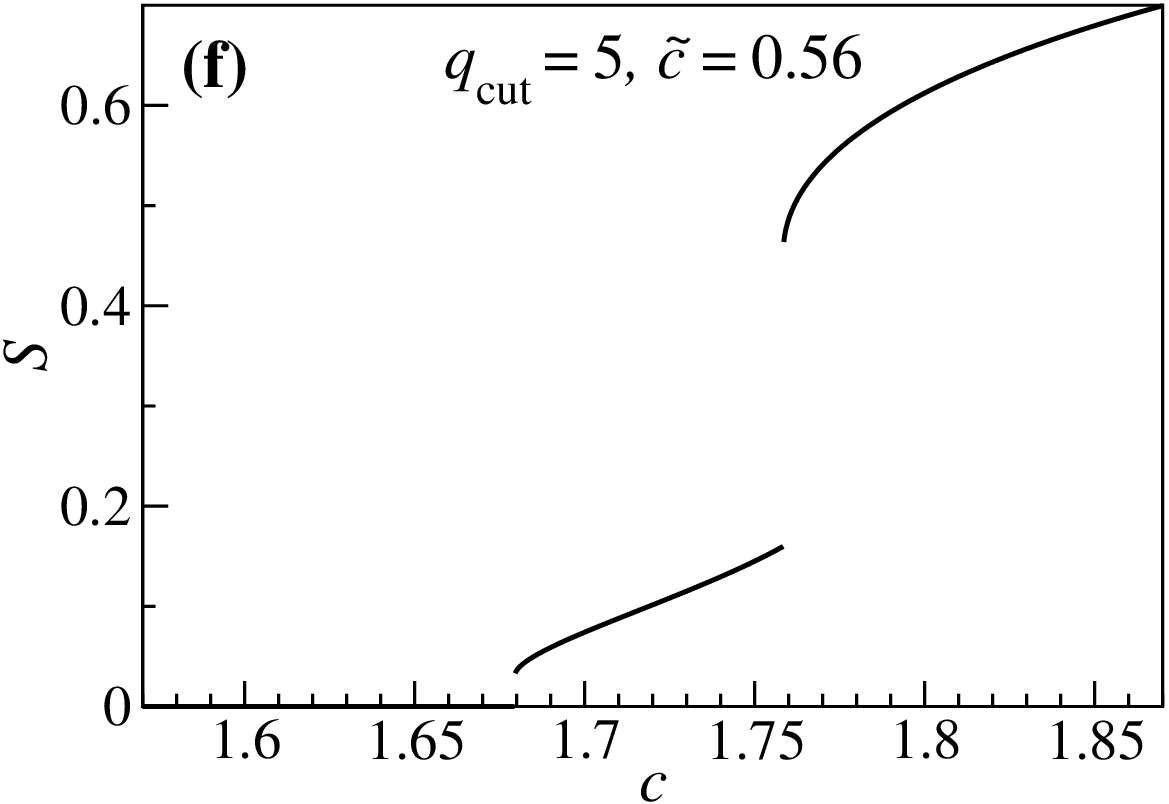}
\end{center}
\caption{(Color online) Phase diagrams in the $\tilde{c}$ vs. $c$ plane for the 
  disassortatively correlated casedefined by Eq. (\ref{60}).
  The appearance of
  the giant mutually connected component with a discontinuous hybrid
  transition is shown as the heavy black line.
(a)--(e) Disassortative correlations of Eq.~(\ref{60}) for 
$q_\text{cut}=10$, $5.7132$, $5.5$, $5$, and $3$. 
For $q_\text{cut}<5.7131...$ instead of a single line of transitions there are two branches, and for $q_\text{cut}<3$ the right-hand side branch extends to $\tilde{c}=\infty$. 
The dashed line is the phase boundary in the uncorrelated case. 
The plots also show the values of the probabilities immediately above the discontinuous transition. 
(f) Multiple transitions of $S$ for $q_\text{cut}=5$, fixed $\tilde{c}=0.56$ and 
$c$ varying along the dotted line of panel (c).
}
\label{f5}
\end{figure*}

An example of 
a symmetric joint degree distribution with disassortative correlations is
\begin{equation}
P(q_1,q_2,\tilde{q})={\cal P}(q_1,c){\cal P}(q_2,c){\cal P}[\tilde{q},A(q_\text{cut}-q_1-q_2)].
\label{60}
\end{equation}
Larger values of $q_\text{cut}$ correspond to weaker correlations, with the anticorrelations becoming stronger as $q_\text{cut}$ decreases.
Once again, the value of $A$ must be chosen to maintain the required
value of $\tilde{c} = \sum_{q_1,q_2,\tilde{q}} \tilde{q}
P(q_1,q_2,\tilde{q}) $,
\begin{equation}
A=\frac{\tilde{c}\left\lfloor q_\text{cut} \right\rfloor !}{(2c)^{1+\left\lfloor q_\text{cut} \right\rfloor}e^{-2c} + (q_\text{cut} - 2c )\Gamma\left( \left\lfloor q_\text{cut} \right\rfloor +1 , 2c\right)}\,,
\label{s4}
\end{equation}
where $ \left\lfloor q_\text{cut} \right\rfloor$ is the largest
integer smaller or equal to $ q_\text{cut}$.

Again using a Poisson distribution
for ${\cal P}(q,c)$  with first moment $c$, and inserting Eq.~(\ref{60}) in Eq.~(\ref{20}) leads to the following system of three transcendental equations:
\begin{widetext}
\begin{eqnarray}
x = && \frac{e^{-2 \tilde{c}}}{\Gamma\left( \left\lfloor q_\text{cut} \right\rfloor \right)} 
\left\{e^{2 \tilde{c}(1-x)}(e^{ \tilde{c} x}-1)^2 \Gamma\left( \left\lfloor q_\text{cut} \right\rfloor \right) 
+
2 e^{\tilde{c}(2-x)}\Gamma\left[ \left\lfloor q_\text{cut} \right\rfloor , \tilde{c}(2-x)\right]
-
e^{2 \tilde{c}(1-x)} \Gamma\left[ \left\lfloor q_\text{cut} \right\rfloor , 2 \tilde{c}(1-x) \right]  \right.
\nonumber
\\
&&
+
 \exp\left[-A(q_\text{cut}-1) (2u+w) + 2 \tilde{c} (1-x) e^{A(2u+w)} \right] \Gamma\left[ \left\lfloor q_\text{cut} \right\rfloor , 2 \tilde{c}(1-x)e^{A(2u+w)} \right]
 \nonumber
\\
&&
 \left.
-
2\exp\left[-A(q_\text{cut}-1) (u+w) + \tilde{c} (2-x) e^{A(u+w)} \right] \Gamma\left[ \left\lfloor q_\text{cut} \right\rfloor , \tilde{c}(2-x)e^{A(u+w)}\right]
\right\}
,
\nonumber
\\[5pt]
u =&&  \frac{1 - w}{2} + \frac{A \exp\left[ -A q_\text{cut} (2u+w) -2\tilde{c} (1-(1-x)e^{A(2u+w)}) \right]}{2 \,\tilde{c}\,  \Gamma\left[ \left\lfloor q_\text{cut} \right\rfloor + 1 \right]}
\left\{ 2  \tilde{c} e^{A(2u+w)}  \left\lfloor q_\text{cut} \right\rfloor (1-x)  \Gamma\left[ \left\lfloor q_\text{cut} \right\rfloor , 2 \tilde{c} (1-x) e^{A(2u+w)} \right] \right.
\nonumber
\\
&&
\left.
-
 q_\text{cut}  \Gamma\left[ \left\lfloor q_\text{cut} \right\rfloor +1 , 2 \tilde{c} (1-x) e^{A(2u+w)} \right]
\right\},
\nonumber
\\[5pt]
w = && 1 - u + \frac{A \exp\left[ - A q_\text{cut} (u+w) - \tilde{c} (2-(2-x)e^{A(u+w)}) \right]}{2 \,\tilde{c}\,  \Gamma\left[ \left\lfloor q_\text{cut} \right\rfloor + 1 \right]}
\left\{  \tilde{c} e^{A(u+w)}  \left\lfloor q_\text{cut} \right\rfloor (2-x)  \Gamma\left[ \left\lfloor q_\text{cut} \right\rfloor , \tilde{c} (2-x) e^{A(u+w)} \right]   \right.
\nonumber
\\
&&
\left.
-
 q_\text{cut}   \Gamma\left[ \left\lfloor q_\text{cut} \right\rfloor +1 , \tilde{c} (2-x) e^{A(u+w)} \right] \right\}
.
\label{s3}
\end{eqnarray}
The expression for the mutual component size $S$ in a network with
these correlations 
is obtained substituting Eq.~(\ref{60}) for $P(q_1,q_2,\tilde{q})$ in Eq.~(\ref{20}):
\begin{eqnarray}
S = && \frac{e^{-2 \tilde{c}}}{\Gamma\left( \left\lfloor q_\text{cut} \right\rfloor {+} 1 \right)} 
\left\{e^{2 \tilde{c}(1-x)}(e^{ \tilde{c} x}-1)^2 \Gamma\left( \left\lfloor q_\text{cut} \right\rfloor  {+} 1 \right) 
+
2 e^{\tilde{c}(2-x)}\Gamma\left[ \left\lfloor q_\text{cut} \right\rfloor  {+} 1, \tilde{c}(2-x)\right]
-
e^{2 \tilde{c}(1-x)} \Gamma\left[ \left\lfloor q_\text{cut} \right\rfloor  {+} 1, 2 \tilde{c}(1-x) \right]  \right.
\nonumber
\\
&&
+
 \exp\left[-A q_\text{cut} (2u+w) + 2 \tilde{c} (1-x) e^{A(2u+w)} \right] \Gamma\left[ \left\lfloor q_\text{cut} \right\rfloor  {+} 1 , 2 \tilde{c}(1-x)e^{A(2u+w)} \right]
 \nonumber
\\
&&
 \left.
-
2\exp\left[-A q_\text{cut} (u+w) + \tilde{c} (2-x) e^{A(u+w)} \right] \Gamma\left[ \left\lfloor q_\text{cut} \right\rfloor  {+} 1, \tilde{c}(2-x)e^{A(u+w)}\right]
\right\}
.
\label{ss3}
\end{eqnarray}
\end{widetext}

Solving these equations, we find a more complex phase diagram than in the positively correlated case. 
Phase diagrams for different values of $q_\text{cut}$ are shown in Fig.~\ref{f5}.

For weak anticorrelations, with $q_\text{cut}$ larger than a specific value $q_\text{cut}^* = 5.7131...$,
panels (a) and (b), the phase diagram is qualitatively similar to the uncorrelated case, 
containing a single line of discontinuous phase transitions with end points $(c,\tilde{c})=(0,1)$ and $(2.4554...,0)$, compare to 
Fig.~\ref{f4}(a). The line of transition moves toward larger values of $\tilde{c}$, while the endpoints again remain fixed.
At $q_\text{cut}{=}q_\text{cut}^*$ a new behavior emerges, as the line of continuous transitions breaks into two branches having different values of $S$ above the transition. 
For $q_\text{cut}{<}q_\text{cut}^*$ the lower branch that starts at $(0,1)$ finishes when it meets the other branch. 
The branch that starts at $(2.4554...,0)$ ends at a finite point for $3{<}q_\text{cut}{<}q_\text{cut}^*$, panels (c) and (d), but extends to $\tilde{c}{=}\infty$ for $q_\text{cut}{\leq}3$, panel (e). 
An example of the solution of these equations is shown in  panel (f)
of Fig.~\ref{f5}. This panel shows the size of the mutual component $S$ along the dotted line in panel (c), compare with the assortative case, Fig.~\ref{f4}(f). 

The disassortative correlations partially separate the nodes 
into two populations: one of nodes with a majority of single
connections, and another with a majority of overlapped connections.
This is reflected in the relative order of the probabilities $x$, $w$,
and  $u$ in panels (b)--(e) of Fig.~\ref{f5}. In the first branch, $u$
(which incorporates effects of overlapped edges) dominates, followed
by $w$ then $x$, whereas in the second branch, $u$ makes the smallest
contribution.
In  the example shown in panel (f) the first jump occurs when a giant
mutual component is first formed, with nodes containing both single
and overlapping edges. Nodes with more than $q_\text{cut}$ single
edges have no overlapped edges. The second jump occurs when a large
number of such nodes are recruited to the giant component.
By comparison, in the uncorrelated and 
assortative cases,
there is no such separation of node populations.

\begin{figure*}
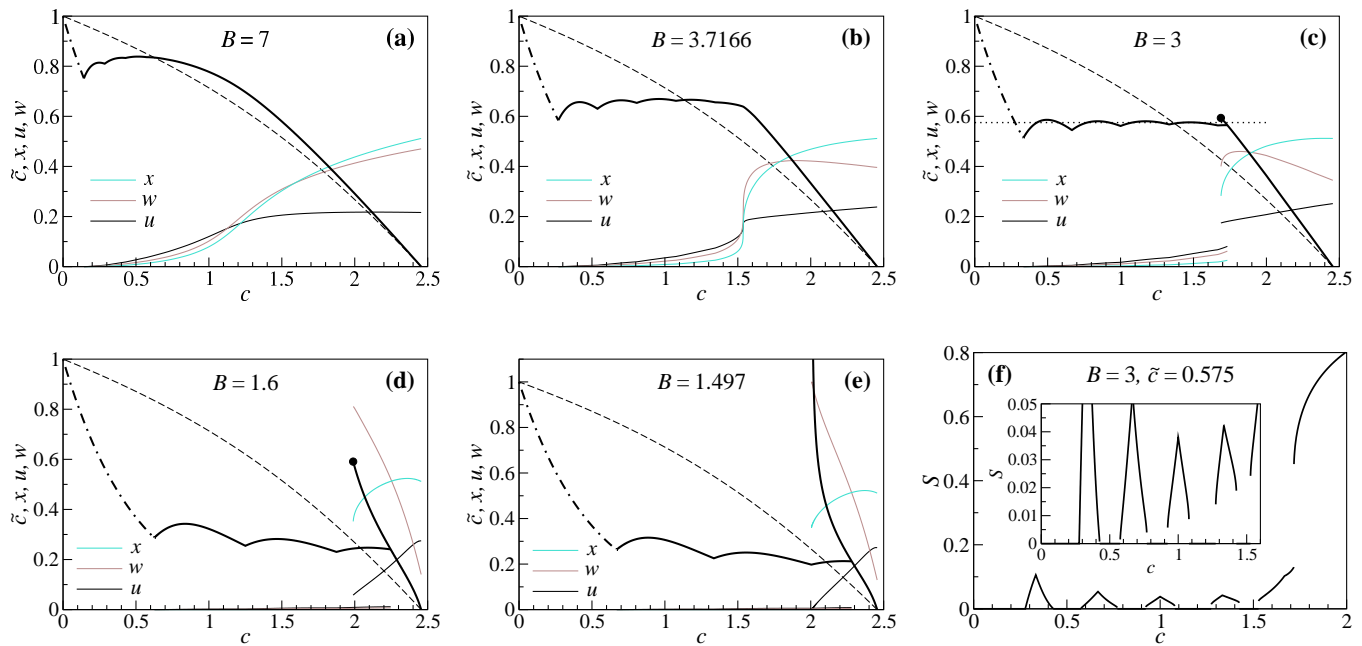

\begin{center}
\includegraphics[width=0.32\textwidth]{fig5a}
\ \ 
\includegraphics[width=0.32\textwidth]{fig5b}
\ \ 
\includegraphics[width=0.32\textwidth]{fig5c}
\\[15pt]
\includegraphics[width=0.32\textwidth]{fig5d}
\ \ 
\includegraphics[width=0.32\textwidth]{fig5e}
\ \ 
\includegraphics[width=0.32\textwidth]{fig5f}
\end{center}
\caption{(Color online) Phase diagrams in the $\tilde{c}$ vs. $c$ plane for the
  disassortatively correlated case defined by Eq. (\ref{s5}).
  The appearance of
  the giant mutually connected component with a discontinuous hybrid
  transition is shown as the heavy black line.
(a)-(e) Disassortative correlations of Eq.~(\ref{60}) with $q_\text{cut}=Bc$, for $B=7$, $3.7166$, $3.5$, $1.6$, and $1.497$.
For $B<3.7165...$ instead of a single line of transitions there are two branches, and for $B<1.497...$ the right-hand side branch extends to $\tilde{c}=\infty$.
The dot-dashed curve is a line of continuous phase transitions that take place at $e^{-2c}$ for $c<1/B$.
 The dashed line is the phase boundary in the uncorrelated case. 
 The plots also show the values of the probabilities immediately above the discontinuous transition. 
(f) Multiple transitions of $S$ for $B=3$, fixed $\tilde{c}=0.575$ and and c varying along the dotted line of panel (c).
}
\label{f6}
\end{figure*}

We also consider an alternative form for the disassortative correlations, where the cut-off degree varies linearly with the average number of single edges $q_\text{cut}=Bc$,
\begin{equation}
P(q_1,q_2,\tilde{q})={\cal P}(q_1,c){\cal P}(q_2,c){\cal P}[\tilde{q},A(Bc-q_1-q_2)],
\label{s5}
\end{equation} 
for different values of $B$. Here the constant $A$ is determined by Eq.~(\ref{s4}) with $Bc$ substituted for $q_\text{cut}$. Similarly, the self-consistency equations and the expression for $S$ are obtained by substituting $Bc$ for $q_\text{cut}$ in Eqs.~(\ref{s3}) and~(\ref{ss3}), respectively. 

Figure~\ref{f6} shows the solution of the model $q_\text{cut}=Bc$ for different $B$, which is qualitatively similar to the one of the model with constant $q_\text{cut}$. We again find a single line of discontinuous hybrid transitions for weak anticorrelations, panels (a) and (b).
As before, at a specific strength of the anticorrelations, $B=3.7165...$, the line of transition breaks into two branches. Below this value of $B$, panels (c) and (d), we find two branches of transitions, with the end point of the second branch diverging to $\tilde{c}=\infty$ for  $B<1.497...$, panel (e).
The different form of the degree correlation function gives a phase boundary with a more complex shape. A straight path in the phase plane may cross this line multiple times, giving multiple hybrid transitions. An example is shown in panel (f), which shows the size of the mutual component $S$ along the dotted line in panel (c). The size $S$ jumps each time a line of discontinuous transitions is crossed. 

As an effect of the correlations of Eq.~(\ref{s5}), when $c\leq 1/B$ only nodes without single edges (i.e. $q_1+q_2=0$) can have overlapping edges, which results in two disjoint subgraphs. On one hand, the subgraph containing all the single edges has a relative size $1-e^{-2c}$, and does not contain a mutual component for $c<2.4554...$. On the other hand, the subgraph consisting of the remaining nodes contains all the overlapping edges and has a relative size $e^{-2c}$. As a result, in the region $c\leq 1/B$ the overlapped subgraph behaves as a single-layer classical random graph, and undergoes a standard continuous transition at the dot-dashed line in Fig.~\ref{f6}. The continuous transition takes place when the average degree in the overlapped subgraph equals $1$, that is, when $\tilde c/e^{-2c}=1$.

\section{Numerical simulations}

\begin{figure}[h!]
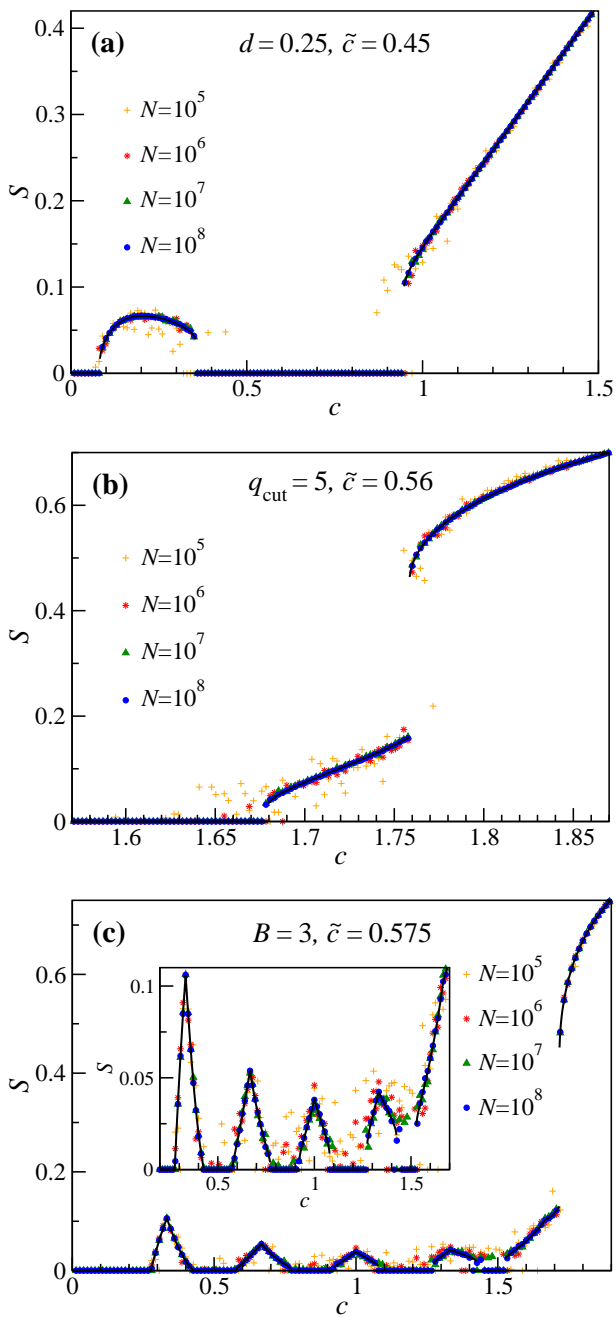

\begin{center}
\includegraphics[width=0.45\textwidth]{fig3f_sim}\\[10pt]
\includegraphics[width=0.45\textwidth]{fig4f_sim}\\[10pt]
\includegraphics[width=0.45\textwidth]{fig5f_sim}
\end{center}
\caption{(Color online) Size of the largest mutually connected component $S$ as a function of mean non-overlapped degree $c$ for correlated two layer multiplex with degree distribution of the form Eq. (\ref{40}). Each point corresponds to a single network realisation of size $N=10^5$ (orange plusses), $10^6$ (red diamonds), $10^7$ (green triangles), or $10^8$ (blue circles). (a) Assortative correlations between overlapped and non-overlapped degree of the form Eq. (\ref{50}). 
(b) Disassortative correlations as in Eq. (\ref{60}). (c)  Disassortative correlations as in Eq. (\ref{s5}). Theoretical results are also plotted (continuous black line), compare Figs. \ref{f4}(f), \ref{f5}(f), and \ref{f6}(f), which show theoretical results for the same parameter choices. 
}
\label{f7}
\end{figure}

In this section we present experimental results obtained from simulations of two-layered systems. In the first stage, we generate networks with the desired properties, namely a given average degree of each type of single edges, $c$, and the particular function $f(x)$ on the correlated joint degree distribution of Eq. (4). Recall that the $f(q_1 + q_2)$ is the average number of overlapping edges of nodes with $q_1$ edges of type 1 and $q_2$ edges of type 2. For a network with $N$ nodes, we first place $cN$ single links of each type connecting pairs of nodes chosen uniformly at random. Finally, we distribute $\tilde{c}N$ overlapping edges, by choosing pairs of nodes where each node gets picked independently with probability proportional to $f(q_1 + q_2)$.

We find the giant mutual component of the resulting network by iteratively following these steps:

(i) We find the largest cluster in each layer, and remove all nodes not belonging simultaneously to the largest cluster in both layers and all edges connected to the removed nodes. (Of course, in this stage of the algorithm, an overlapping edge is treated as two independent single edges.)

(ii) If the remaining subset of nodes and edges is the same size as before executing step (i), i.e., nothing was removed, then it is the giant mutual component. Otherwise, we return to step (i), but only with the remaining of the system.

This algorithm stops when all of the remaining nodes are in the same cluster in both layers, which means every node in this cluster can reach every other by paths strictly inside the cluster. This subset of nodes and edges forms the mutually connected cluster. 

We performed simulations according to this prescription for systems of size $N=10^5$, $10^6$, $10^7$, and $10^8$ 
for each of the three example cases described in Sect.~\ref{correlated}. For each value of $c$ we generated a network of a given size and calculated the size of the largest cluster, according to the method described above.
The results of the simulations are presented in Fig. \ref{f7}, and show an excellent agreement with our theoretical predictions. Fluctuations reduce with system size, and are only significant for the case $N=10^5$.

\section{Conclusions}\label{conclusions}

In this Paper we have developed a theory for 
locally tree-like multiplex networks with overlapped edges 
allowing the giant mutual component in these networks to be found. 
The simplicity of our theory enables the study of more difficult and
rich cases than previously possible.
In particular this method allows for arbitrary interlayer degree
correlations and correlations in overlapped and single edge degrees.
These correlations 
qualitatively change the phase diagrams for multiplex networks.
We found qualitatively new features: new
phase diagrams with multiple and recursive hybrid phase transitions.
We
observed the new phase diagrams for a particular form of the
correlation function.
To confirm our observations, we 
also considered a different form of correlations than in
Eq.~(\ref{60}) and arrived at 
similar results (see Supplementary Material). This allows us to
suggest that our qualitative findings are valid for a wide range of
correlations.
Our method follows logically from the structure of the problem, writing an equation for each possible way to encounter the giant mutual component. Further generalisations, such as addition of more layers, should be able to be treated using the same logic, requiring more equations but not new techniques. Indeed since this paper was first submitted, such a generalization has already been proposed \cite{cellai2016message}.
Previous methods do not have this advantage.
Overlapping layers are an unavoidable feature of interdependent
networks, naturally emerging in various problems
\cite{azimi2015cooperative}, yet they make theoretical treatment
much more difficult. 
We suggest that the simplicity and tractability of our theory and
results will make this task much easier.

\begin{acknowledgments}
This work was partially supported by the FET
proactive IP project MULTIPLEX 317532. GJB was supported by
the FCT grant No. SFRH/BPD/74040/2010.
\end{acknowledgments}

\bibliography{bibliography}



\appendix*

\section{Extended form of Main Equations}
To aid the reader, we give here forms of the equations for $x,y,u,v,$ and $w$ with terms corresponding to each term of the diagrammatic equations Fig. \ref{f2}. Simplification of these equations leads to Eqs. (\ref{20}).
\begin{widetext}
\begin{eqnarray}
x&=&\sum_{q_1,q_2,\tilde{q}}\frac{q_1}{\avg{q_1}}P(q_1,q_2,\tilde{q})\left\{\left[1-(1{-}x)^{q_1-1}\right]\left[1-(1{-}y)^{q_2}\right](1{-}u{-}v{-}w)^{\tilde{q}}\right.
\nonumber 
\\
&&+\left[1-(1{-}x)^{q_1-1}\right]\left[(1{-}u{-}w)^{\tilde{q}}-(1{-}u{-}v{-}w)^{\tilde{q}}\right]
+\left[1-(1{-}y)^{q_2}\right]\left[(1{-}v{-}w)^{\tilde{q}}-(1{-}u{-}v{-}w)^{\tilde{q}}\right]
\nonumber 
\\ [5pt]
&& \left.+\left[(1{-}w)^{\tilde{q}}-(1{-}w{-}u)^{\tilde{q}}-(1{-}w{-}v)^{\tilde{q}}+(1{-}w{-}u{-}v)^{\tilde{q}}\right]+\left[1-(1{-}w)^{\tilde{q}}\right]\right\}
,
\nonumber 
\\[5pt]
y&=&\sum_{q_1,q_2,\tilde{q}}\frac{q_2}{\avg{q_2}}P(q_1,q_2,\tilde{q})\left\{\left[1-(1{-}x)^{q_1}\right]\left[1-(1{-}y)^{q_2-1}\right](1{-}u{-}v{-}w)^{\tilde{q}}\right.
\nonumber 
\\
&&+\left[1-(1{-}x)^{q_1}\right]\left[(1{-}u{-}w)^{\tilde{q}}-(1{-}u{-}v{-}w)^{\tilde{q}}\right]
+\left[1-(1{-}y)^{q_2-1}\right]\left[(1{-}v{-}w)^{\tilde{q}}-(1{-}u{-}v{-}w)^{\tilde{q}}\right]
\nonumber 
\\[5pt]
&&\left.+\left[(1{-}w)^{\tilde{q}}-(1{-}u{-}w)^{\tilde{q}}-(1{-}v{-}w)^{\tilde{q}}+(1{-}u{-}v{-}w)^{\tilde{q}}\right]+\left[1-(1{-}w)^{\tilde{q}}\right]\right\}
,
\nonumber
\\[5pt]
u&=&\sum_{q_1,q_2,\tilde{q}}\frac{\tilde{q}}{\avg{\tilde{q}}}P(q_1,q_2,\tilde{q})\left\{\left[1-(1{-}x)^{q_1}\right](1{-}y)^{q_2}(1{-}u{-}v{-}w)^{\tilde{q}-1}+
(1{-}y)^{q_2}\left[(1{-}v{-}w)^{\tilde{q}-1}-(1{-}u{-}v{-}w)^{\tilde{q}-1}\right]\right\}
,
\nonumber 
\\[5pt]
v&=&\sum_{q_1,q_2,\tilde{q}}\frac{\tilde{q}}{\avg{\tilde{q}}}P(q_1,q_2,\tilde{q})\left\{(1{-}x)^{q_1}\left[1-(1{-}y)^{q_2}\right](1{-}u{-}v{-}w)^{\tilde{q}-1}
+(1{-}x)^{q_1}\left[(1{-}u{-}w)^{\tilde{q}-1}-(1{-}u{-}v{-}w)^{\tilde{q}-1}\right]\right\}
,
\nonumber 
\\[5pt]
w&=&\sum_{q_1,q_2,\tilde{q}}\frac{\tilde{q}}{\avg{\tilde{q}}}P(q_1,q_2,\tilde{q})\left\{\left[1-(1{-}x)^{q_1}\right]\left[1-(1{-}y)^{q_2}\right](1{-}u{-}v{-}w)^{\tilde{q}-1}\right. 
\nonumber 
\\
&&+\left[1-(1{-}y)^{q_2}\right]\left[(1{-}v{-}w)^{\tilde{q}-1}-(1{-}v{-}w{-}u)^{\tilde{q}-1}\right]+\left[1-(1{-}x)^{q_1}\right]\left[(1{-}u{-}w)^{\tilde{q}-1}-(1{-}v{-}w{-}u)^{\tilde{q}-1}\right]
\nonumber 
\\[5pt]
&&\left.+\left[(1{-}w)^{\tilde{q}-1}-(1{-}w{-}u)^{\tilde{q}-1}-(1{-}w{-}v)^{\tilde{q}-1}+(1{-}w{-}u{-}v)^{\tilde{q}-1}\right]+\left[1-(1{-}w)^{\tilde{q}-1}\right]\right\}
.
\end{eqnarray}
\end{widetext}
\end{document}